\documentclass[pra, twocolumn,english,showpacs,floatfix]{revtex4}
\usepackage{graphics} \usepackage{graphicx} \usepackage{epsfig}
\usepackage{amssymb} \usepackage{amsmath}

\begin{document}
\renewcommand{\i}{\ensuremath{ \textrm{i}}}
\renewcommand{\Re}{\ensuremath{ \textrm{Re}}}
\newcommand{\rperp}{\ensuremath{ r_\perp}}
\newcommand{\rpara}{\ensuremath{ r_\parallel}} 
\newcommand{\etal}{{\it et al.}}  \newcommand{\ket}[1]{|{#1}\rangle}
\newcommand{\bra}[1]{\langle{#1}|}
\newcommand{\braket}[1]{\langle{#1}\rangle} \newcommand{\ad}{a^\dagger}
\newcommand{\e}{\ensuremath{\mathrm{e}}}
\newcommand{\norm}[1]{\ensuremath{| #1 |}}
\newcommand{\aver}[1]{\ensuremath{\big<#1 \big>}}
\newcommand{\qperp}{\ensuremath{q_\perp}} \newcommand{\qpara}{\ensuremath{q_\parallel}}

\title{Dipolar bosons in a planar array of one-dimensional tubes}

\author{C. Kollath$^{1,2}$, Julia S. Meyer$^{3}$, T. Giamarchi$^{1}$}

\affiliation{$^1$DPMC-MaNEP, University of Geneva, 24 Quai
  Ernest-Ansermet,
  1211 Geneva, Switzerland\\$^2$Centre de Physique Th\'eorique, Ecole Polytechnique, 91128
 Palaiseau Cedex, France \\
  $^3$Department of Physics, The Ohio State University, Columbus, Ohio
  43210, USA}

\begin{abstract}
  We investigate bosonic atoms or molecules interacting via dipolar
  interactions in a planar array of one-dimensional tubes. We consider the
  situation in which the dipoles are oriented perpendicular to the tubes
  by an external field. We find various quantum phases reaching from a
  `sliding Luttinger liquid' phase to a two-dimensional charge density wave ordered phase. Two
  different kinds of charge density wave order occur: a stripe phase in
  which the bosons in different tubes are aligned and a checkerboard
  phase. We further point out how to distinguish the occurring phases
  experimentally.

\end{abstract}

\maketitle

Recent experiments in ultracold atoms have achieved the realization of
various quantum phases. These phases reach from the superfluid and
Mott-insulator of bosons in optical lattices to a BCS phase and a Bose-Einstein condensate of molecules in fermionic
gases~\cite{BlochZwerger2007}. In addition, it has been possible to
engineer different trapping geometries which enabled the study of the
remarkable physics of one-dimensional quantum
systems~\cite{StoeferleEsslinger2004,ParedesBloch2004,KinoshitaWeiss2004}.
In such systems, known as Luttinger liquids~\cite{Giamarchibook}, the
interactions play a major role and lead to properties quite different from
their higher-dimensional counterparts. This has been nicely demonstrated
investigating the crossover between an array of decoupled one-dimensional
tubes and a quasi-three-dimensional system by lowering the potential
barrier between the tubes~\cite{StoeferleEsslinger2004}.  Since the
interactions in most cold atomic gases have a short range of the order of
a few nm and the distance between the tubes is typically $~$500 nm, the
only coupling between the tubes is generally provided by the hopping of
particles between tubes. For bosons this leads to a dimensional crossover
between the peculiar one-dimensional phases and a quasi-three-dimensional
superfluid~\cite{HoGiamarchi2004}.

However, other fascinating phases may occur when one-dimensional tubes are
directly coupled by interactions.  These include in particular the so
called `sliding Luttinger liquid'
(SLL)~\cite{Schulz1983,EmeryLubensky2000,VishwanathCarpentier2001} that
was studied in connection with high-Tc superconductors and stripe
physics~\cite{KivelsonHowald2003} for fermionic systems. In the SLL phase
the typical properties of one-dimensional systems, namely algebraically
decaying correlations, survive despite the coupling. Thus the Fermi liquid
phase which usually occurs in more than one dimension is suppressed.
Experimentally the SLL has not been observed yet.

In cold atomic gases these phases could not be explored so far due to the
lack of interactions extending over the range of the inter-tube distances.
However the experimental progress in the realization of quantum degenerate
atomic and molecular gases with dominating
dipole-dipole interactions~\cite{LahayePfau2007} have shown the potential to bridge that gap.
Particularly promising are polar molecules with large electric dipole
moments~\cite{BethlemMeijer2000,WangStwalley2004,MeerakkerMeijer2005,RiegerRempe2005,SageDeMiller2005},
since the strength of the dipolar interaction is considerable over the
range of tube spacings~\cite{DamskiZoller2003}.

\begin{figure} [ht]
  \begin{center}
    {\epsfig{figure=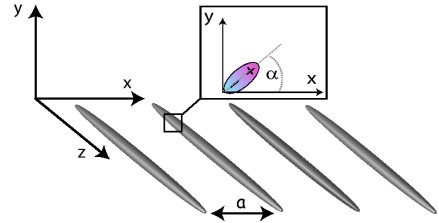,width=0.65\linewidth}}
  \end{center}
  \caption{A planar array of one-dimensional tubes. The dipoles point
    perpendicular to the tube direction. In the inset the angle $\alpha$
    of the orientation of the dipoles with respect to the plane of the
    array is shown.  }
  \label{fig:array}
\end{figure}

In this manuscript we investigate the possible quantum phases of a dipolar
bosonic gas in a planar array of one-dimensional homogeneous tubes
(Fig.~\ref{fig:array}). This situation can be
achieved experimentally by using a strong two-dimensional
optical lattice. If sufficiently strong, this
lattice suppresses the hopping of particles between different tubes. Even in this situation the dipole-dipole interaction couples the tubes. The
dipole-dipole interaction is attractive or repulsive depending on the
relative orientation of the dipoles. This anisotropy causes interesting phenomena
such as an
instability towards collapse~\cite{SantosLewenstein2000,GoralPfau2000}.
Here we focus on the situation where the dipoles are
aligned by an additional external field. The orientation is chosen perpendicular to the direction of the
tubes, since this configuration is the most stable \footnote{The
  stability of a dipolar gas has been studied for a pancake
  structure~\cite{BuechlerZoller2007}. The discussion can analogously be transferred to
  the one-dimensional tubes.}. The orientation with respect to
the plane of the array is varied.

\begin{figure} [ht]
  \begin{center}
    {\epsfig{figure=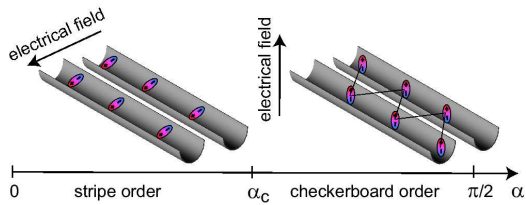,width=0.8\linewidth}}
  \end{center}
  \caption{ A sketch of the transition between a stripe and checkerboard CDW order depending on
    the orientation of the dipoles is shown. 
  }
  \label{fig:cdwsketch}
\end{figure}

Assuming vanishing hopping between the tubes we find four different
regimes
: a SLL state with (i)
dominating superfluid correlations or (ii) dominating charge density wave (CDW)
correlations along the tube forms, (iii) checkerboard or stripe CDW ordering develops
(cf.~Fig.~\ref{fig:cdwsketch}), and (iv) an instability towards collapse
occurs \footnote{Note, that in an actual system with finite radial
  extension of the tubes, the instability may shift to lower values of the
  dipolar
  interaction~\cite{BuechlerZoller2007}.}. The finding of a SLL phase is novel in a bosonic system. The SLL has
power-law correlation functions along the tubes which can be dominating
superfluid correlations (i) or dominating CDW correlations (ii). The two
regimes are connected by a crossover.  We discuss how cold quantum gases
could provide the opportunity to observe the SLL phase for the first time
experimentally.

Interacting bosons in an array of
one-dimensional tubes (cf.~Fig.~\ref{fig:array}) can be described by the
Hamiltonian
\begin{eqnarray}
\label{eq:Hor}
\!\!\!\!\!\!H&\!\!=\!\!&\sum_{j}\int \! {\textrm d}{z} \left ( \frac{1}{2M}\norm{\partial_z \hat{\Psi}_j(z)}^2 + \frac{g}{2} \hat{\rho}_j(z) \hat{\rho}_{j}(z)\right) \nonumber\\
&&+ \sum_{j, j'} \int \!{\textrm d}{z}\, {\textrm
    d}{z'}\; V_d((j\!-\!j')a,0,z\!-\!z') \hat{\rho}_j(z) \hat{\rho}_{j'}(z').
\end{eqnarray}
Here $\hat{\Psi}_j$ is the bosonic anihiliation operator in tube $j$ and
$\hat{\rho}_j=\hat{\Psi}_j^\dagger \hat{\Psi}_j$ is the density operator. We use $\hbar =1$. The tube distance is $a$, $M$ is the mass of the particles, and $g$ is the
strength of the $\delta$-interaction between the particles.

The last term in Hamiltonian (\ref{eq:Hor}) stems from the dipole-dipole
interaction. Its amplitude is $V_d(\mathbf{r})= V_0 (\mathbf{ \hat{d}}^2- 3
(\mathbf{\hat{ d}}\mathbf{
    \hat{r}})^2)/{r^3}$, where $\mathbf{\hat{d}} 
  = (\cos\alpha, \sin \alpha, 0)$ denotes the direction of the dipole
  moment and $\mathbf{ \hat{r}}$ is the unit vector. The interaction
  strength $V_0$ is given by $V_0=d^2\mu_0/(4\pi)$ for magnetic and
  $V_0=d^2/(4\pi\epsilon_0)$ for electric dipoles.  Here $d$ is the
  strength of the dipole moment, and $\mu_0$ ($\epsilon_0$) is the vacuum
  permeability (permittivity). 

The low-energy properties of the system can be described using the
bosonization approach~\cite{Schulz1983,Giamarchibook}. Two bosonic fields,
$\hat{\phi}_j$ and $\hat{\theta}_j$, are
introduced to describe the modes of the system, where $\hat{\phi}_j$ and
$\hat{\theta}_j$ are related to the amplitude and phase of the operator
$\hat{\Psi}_j$, respectively. The Hamiltonian in the absence of the dipolar
interaction becomes
\begin{eqnarray}
\label{eq:H}
H_0&\!\!=\!\!&\frac u{2\pi}\sum_{j} \int \textrm{d}z  \left( K 
 (\nabla \hat{\theta}_{j}(z))^2 + \frac{1}{K}  (\nabla \hat{\phi}_{j}(z))^2 \right)\!,
\end{eqnarray}
where the parameters
$u$, the velocity, and $K$, the Luttinger parameter, depend on the
microscopic details of the underlying system~\cite{Giamarchibook}.
For weak interaction the relations $K \approx \pi\sqrt{\frac{\rho_0}{Mg}}$ and $u \approx \sqrt{\frac{g\rho_0}{M}}$ hold. Here $\rho_0$ is the average density of bosons.

The dipolar interaction can be expressed in the bosonization language using
$\hat{\rho}_j(z)= \rho_0 -(1/\pi) \nabla \hat{\phi}_j(z)+\rho_0 \sum_{p\not = 0}
\e^{\i 2p(\pi\rho_0 z-\hat{\phi}_j(z))}$, where $p$ is an integer. The smooth part of the density $\sim
\nabla\hat{\phi}_j$ yields a non-local quadratic term in the Hamiltonian whereas
the first harmonic $|p|=1$ introduces backscattering terms. Neglecting
higher harmonics, $\norm{p}>1$, the contribution to the Hamiltonian of the dipolar interaction thus becomes
\begin{eqnarray}
H_d&=&\frac{1}{\pi^2}\sum_{j,n}\int {\textrm d}{z}\, {\textrm
    d}{z'}\; V_d(na,0,z\!-\!z')\times\nonumber\\
&&\quad\quad\quad\times\Big(\nabla
    \hat{\phi}_{j}(z)\nabla \hat{\phi}_{j+n}(z')+  \hat{O}^b_{j,n}(z,z')\Big),
\label{eq:Hd}
\end{eqnarray}
where $\hat{O}^b_{j,n}(z,z')\propto \cos
(2\hat{\phi}_j(z)-2\hat{\phi}_{j+n}(z'))$ are the backward scattering operators. 

Assuming that the average particle
  spacing is large compared to the size of the dipoles, the detailed structure of the interaction at short
  distances can be neglected~\cite{YiYou2001,MarinescuYou1998}, and any
  possible further contribution to the interaction on a length scale below
  a cutoff $1/\rho_0$ can be absorbed into the values of $u$ and $K$.  By this
  definition $K$ can now range from very large values for a weakly
  interacting gas without dipolar interactions to values smaller than one
  in the presence of finite dipolar interactions as shown in
  Ref.~\cite{CitroChiofalo2007} for a purely dipolar one-dimensional
  system. Reaching values of $K$ smaller than one in a bosonic system is a
  characteristic of non-local interactions \cite{Giamarchibook}. In the following we use
  the dimensionless ratio of the dipolar interactions and the $s$-wave
  scattering given by $\gamma=4 V_0 K/(a^2 u )$.



\begin{figure*} [t]
  \begin{center}
    {\epsfig{figure=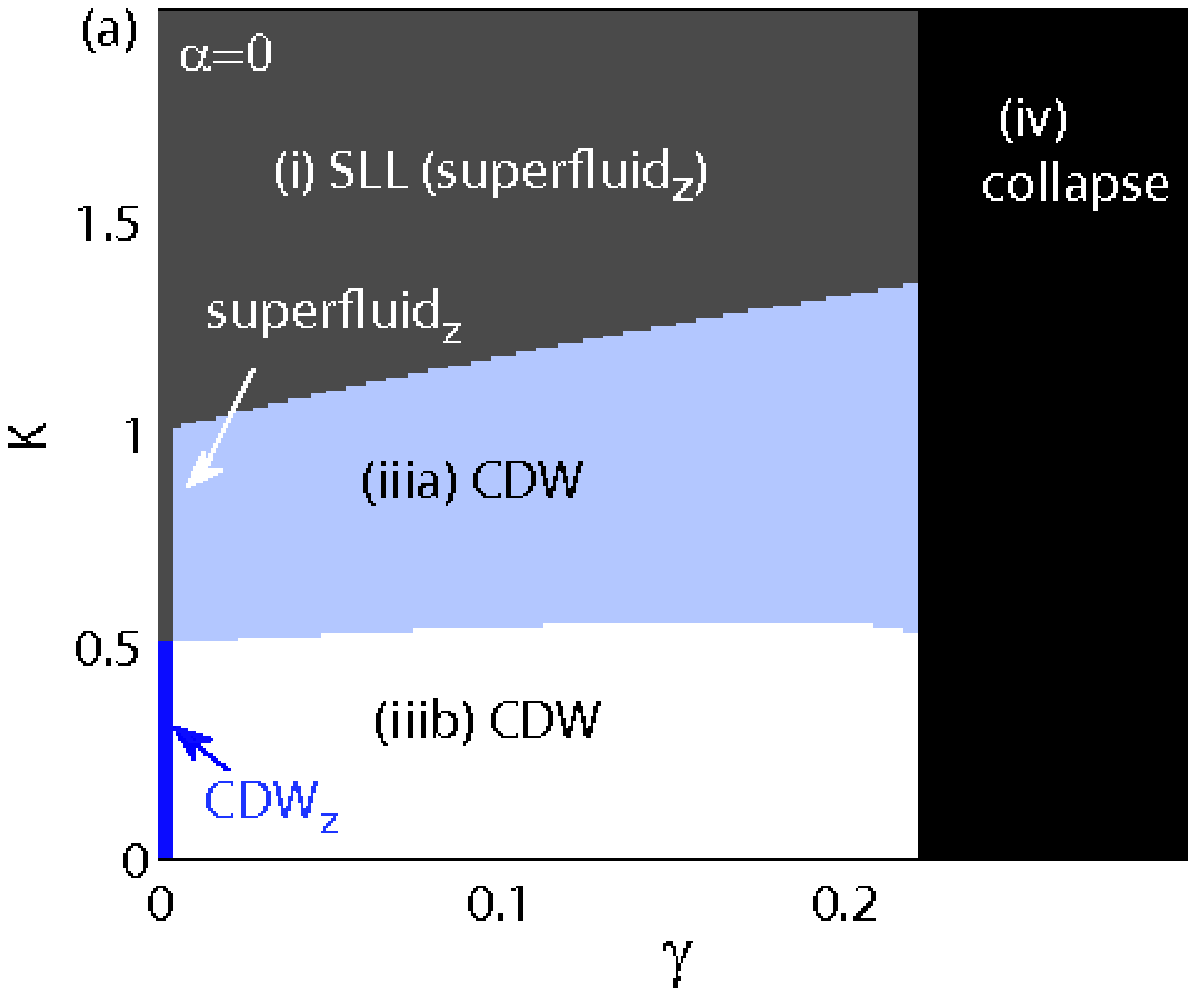,width=0.35\linewidth}}\qquad\qquad
    {\epsfig{figure=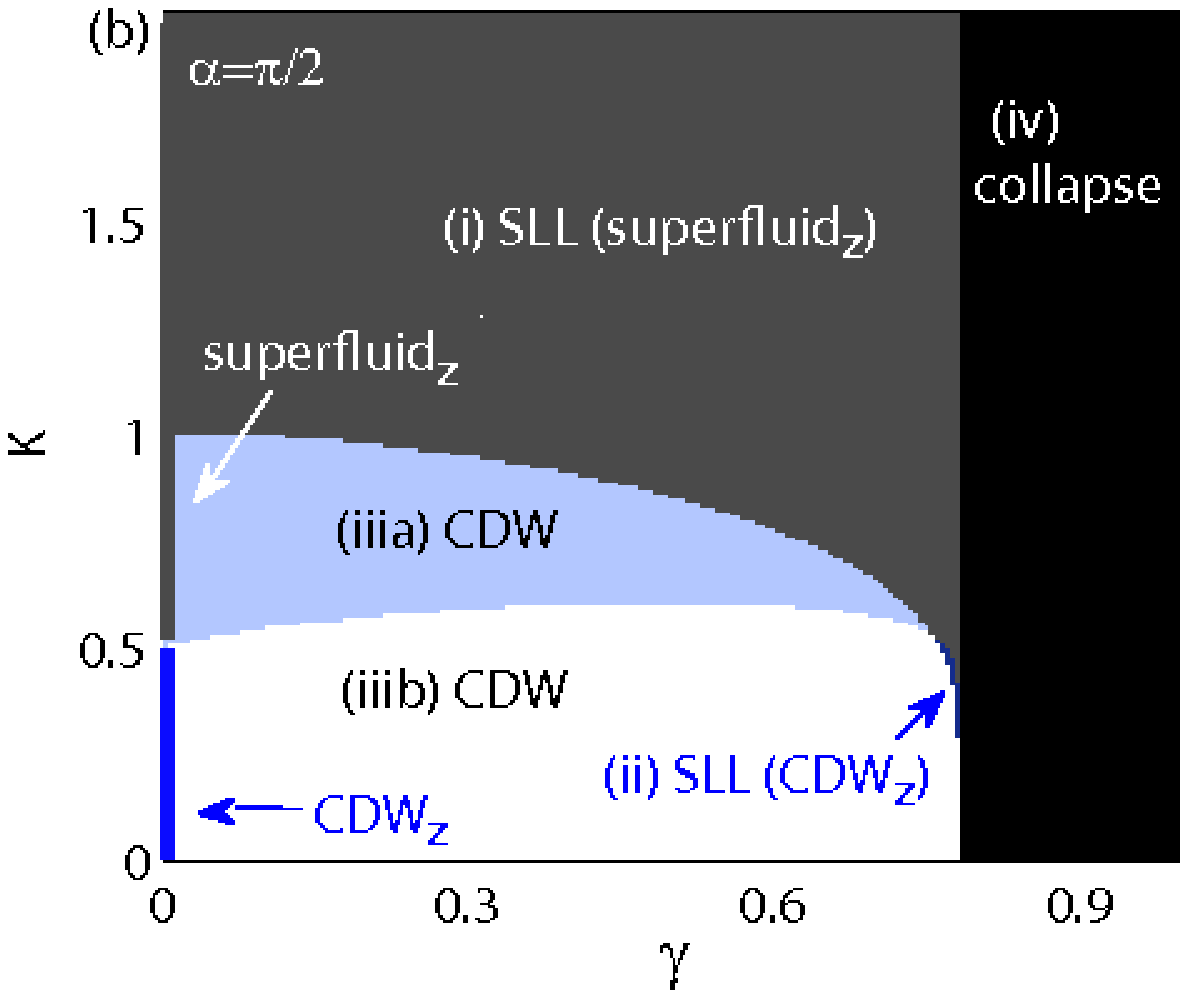,width=0.35\linewidth}}
  \end{center}
  \caption{(Color online) Different quantum phases occuring for the directions (a) $\alpha=0$ and
    (b) $\alpha=\pi/2$.  Experimentally $K$ can be changed varying the $s$-wave
    scattering length of the particles. Further tuning the dipolar interaction
    would correspond to changing both $\gamma$ and $K$. The
    subscript `$z$' denotes the dominant correlations along the tube. 
    }
  \label{fig:rel}
\end{figure*}



In order to determine the quantum phases occurring in the system, we investigate the relevance of the superfluid and CDW operators along and perpendicular to the tubes. 
To find the dominant
correlations we first consider the effects of the quadratic part of $H_0+H_d$ by comparing the relevance of the density-density
correlations $\aver{\hat{\rho}_j(z) \hat{\rho}_j(0)} \propto (1/z)^{2K
  G_{\phi,\parallel}}$ and the single particle correlations
$\aver{\hat{\Psi}_j^\dagger(z) \hat{\Psi}_j(0)} \propto (1/z)
^{G_{\theta,\parallel}/(2K)}$. The algebraic decay of both
correlations is typical for a Luttinger liquid~\cite{Giamarchibook}. We find that the
exponents depend on the strength of the dipolar interaction and the
orientation of the dipoles through $G_{\phi,\parallel}(\gamma, \alpha)=
1/(2\pi)\sum_{\qperp}(1+\gamma \tilde{V}(\qperp,0))^{-1/2} $ and
$G_{\theta,\parallel}(\gamma,\alpha)= 1/(2\pi)\sum_{\qperp} (1+\gamma
\tilde{V}(\qperp,0))^{1/2}$. Here
$\tilde{V}(\qperp,\qpara=0)=2(a\rho_0)^2-4\cos(2\alpha) \Re[{\rm
  Li}_2(\e^{\i a\qperp})]$, where $\tilde{V}(\qperp,
  \qpara)=V_d(\qperp, \qpara)/(4\pi V_0) $ denotes the dimensionless interaction
  strength, ${\rm Li}_2$ the polylogarithm, and
$\Re$ the real part. $\qperp$ and $\qpara$ are the
  components of the momentum perpendicular and parallel to
the tubes, respectively. To perform the calculations a discrete Fourier transform is taken in the
$x$-direction perpendicular to the tubes, and a continuous Fourier transform along the
  tubes. We further approximate the momentum dependence of the dipolar
interaction along the tubes by the low-momentum value at $\qpara=0$ which
yields the main contribution to the long-distance behavior of the
correlation functions. This corresponds to using an effective interaction
that is local along the tube and whose strength is given by the dipolar
interaction integrated over the tube~\footnote{In contrast to the
  three-dimensional case, in one dimension the integral over $1/r^3$ is infrared finite.}.

Comparing the obtained exponents, the CDW correlations in the tube dominate at long
distances if
$$
2KG_{\phi,\parallel}- \frac{1}{2K}G_{\theta,\parallel}<0.$$
For
decoupled tubes the functions reduce to $G_{\phi,\parallel}=1$ and
$G_{\theta,\parallel}=1$. In this case one recovers the crossover for a
single tube~\cite{Giamarchibook, CitroChiofalo2007} between a region in
which the superfluid correlations are dominant for $K>1/2$ and a region in
which the CDW correlations are dominant for $K<1/2$ (cf.~$\gamma=0$ line
in Fig.~\ref{fig:rel}).

For coupled tubes the inter-tube operators can become relevant and change
the nature of the occurring quantum phases. If the pre-factors of the
operators $ \hat{O}^b_{j,n}$ are small, their relevance can be determined by
looking at their scaling dimension. The scaling dimension of the operators $\hat{ O}^b_{j,n}$ is
$2K G_{\phi,\perp,n}$ using $G_{\phi,\perp,n}= 1/(2\pi) \sum_{\qperp}
(1+\gamma \tilde{V}(\qperp,0))^{-1/2} \left(1-\cos(na\qperp)\right)$.
Therefore a backscattering operator between tubes at distance $na$ becomes
relevant and can induce charge ordering, if $$2KG_{\phi,\perp,n}-2 <0.$$

In Fig.~\ref{fig:rel} the phase diagram depending on $\gamma$ and $K$ is shown for two different orientations of the dipoles \footnote{Only the relevance of the intra-tube operators as
    well as the backscattering operators between neighboring tubes
    ($n=1$) are represented. The backscattering operators $\hat O^b_{j,n}$ with
    $n\geq 2$ are less relevant except for small parameter regimes at
    large $\gamma$ (checked up to $n=5$).}. Experimentally $K$ can be varied by
adjusting the short-range interaction between the particles. Varying the
dipolar interaction strength changes both, $\gamma$ and $K$.  In
Fig.~\ref{fig:rel}(a) the dipoles are pointing along the direction of the
array ($\alpha=0$) leading to an attractive inter-tube interaction at
short distances. By contrast in Fig.~\ref{fig:rel}(b) the dipoles are
oriented perpendicular to the plane of the array ($\alpha=\pi/2$)
resulting in a purely repulsive inter-tube interaction. The
phase boundaries are shown for the case where the spacing between
tubes equals the average particle distance, i.e.,
$a\rho_0=1$~\footnote{The phase boundaries may shift due to a
  renormalization of the parameters. A renormalization procedure can be
  performed using the approximation of an effective short range
  interaction. However, due to the approximation this would not yield more
  information about the phase transitions in terms of the experimental
  parameters.}.

For $\gamma\ll 1$ the behavior for the different orientations of the
dipoles is very similar: For $K \gtrsim 1$ a SLL with dominant superfluid correlations (regime (i))
occurs. In contrast to an array of independent tubes, the presence of the
dipolar interaction leads to a coupling of the densities in different
tubes on long length
scales. In the bosonization language the coupling is given by the forward
scattering terms, i.e.,~in Hamiltonian (\ref{eq:Hd}) the terms which
contain $\nabla\hat{\phi}_j(z)\nabla\hat{\phi}_{j'}(z') $ with $j\neq j'$.

The CDW correlations along the tubes become relevant for
values $K\lesssim 1/2$. However, the CDW correlations
perpendicular to the tubes become relevant already for larger values
$K\lesssim 1$. In the regime $1/2\lesssim K \lesssim 1$ (regime (iiia)), a
recalculation of the intra-tube correlations taking into account the
ordering in the perpendicular direction then leads to a CDW ordered phase
in all directions. In this case, the order in the tube will be weaker than
the order in the direction perpendicular to the tubes. In contrast, for
$K\lesssim 1/2$, the order along the tubes is approximately as strong as
the order perpendicular to the tubes, since both the CDW ordering along
and perpendicular to the tubes are relevant (regime (iiib)).

Above a critical value of $\gamma$ an instability towards collapse (regime
(iv)) occurs. The instability occurs if there exists a $\qperp$ for which $1+\gamma \tilde{V}(\qperp) <0$. The physics of the instability is different for the shown cases $\alpha=0$ and $\alpha=\pi/2$. For $\alpha=0$ the attractive inter-tube interaction overcomes the repulsive intra-tube interation and causes the collapse of the particles inside the tubes and an alignement of these perpendicular to the tubes. However, the instability for $\alpha=\pi/2$ stems from a strong repulsive inter-tube interaction which dominates over the intra-tube interaction. This can occur, if the tubes are very close and the contact interaction is attractive. Here the particles collapse inside a tube in order to avoid the interaction with the particles in the neighbouring tubes.

For intermediate values of $\gamma$ the behavior depends on the
orientation of the dipoles. For $\alpha=0$ the CDW order can already be reached for
values of $K>1$, i.e., the dipolar interaction enhances the tendency of
the system to order. By contrast for $\alpha =\pi/2$, a larger value of $\gamma$ can
destabilize the CDW ordering. Thereby a transition between a SLL and a CDW
ordered phase seems to be possible simply by varying the orientation of
the dipoles.  Further for $\alpha =\pi/2$ a small region of SLL with
dominating CDW order (regime (ii)) can be seen in Fig.~\ref{fig:rel}(b) at
large $\gamma$. Whether this survives for realistic experimental
parameters is an open question.

In phase (iii), the form of the CDW order which occurs depends on the
direction of the dipoles (Fig.~\ref{fig:cdwsketch}). In particular if the
dipoles lie in the plane of the tubes, the interaction between tubes is
attractive at short distances and therefore the CDW order of different
tubes is aligned \cite{WangDemler2006}.  By contrast, if the dipoles are perpendicular to the
plane, the interaction between tubes is repulsive and a checkerboard
ordering is expected.  To determine which form of CDW order the system
takes for a given angle $\alpha$ we perform a minimization of the energy
of the system assuming CDW order along the tubes, i.e., we take the
density in tube $j$ to be of the form $\hat{\rho}_j(z)= \rho_0 \left(1+\cos(2\pi
  \rho_0 z -\bar \phi_j)\right)$. Here $\bar \phi_j$ describes the average
phase in tube $j$.  With this ansatz (and using $\delta z= z-z'$) the expression to be minimized
reduces to $$\sum_{j\not = j'} \int d\delta z\; V_d((j\!-\!j')a,0,\delta z)
\cos(2 \pi\rho_0 \delta z - 2 (\bar \phi_j -\bar \phi_{j'})). $$ As
expected, we find a transition between the stripe order for small
$\alpha$ and the checkerboard order for $\alpha$ close to $\pi/2$ (see Fig.~\ref{fig:cdwsketch}).  Assuming only coupling
between nearest-neighbor tubes the transition takes place at
$\cos^2(\alpha_c)= K_1(2 \pi a\rho_0)/\left(2 \pi a\rho_0 K_2(2 \pi
  a\rho_0)\right)$, where $K_i$ are modified Bessel functions.  At the
transition point the tubes experience only a weak coupling to other tubes,
and the correlations in the tube are superfluid or CDW dominated depending
on the parameter regime. Taking the full dipolar interaction into account
we determine the transition point numerically \footnote{Hereby we consider
  two different cases: (i) up to four coupled tubes with arbitrary $\bar
  \phi_j$, and (ii) a large number of tubes, but with a constant phase
  shift $\phi_d=\bar \phi_j-\bar \phi_{j+1}$.}.
The result can hardly be distinguished from the transition point found for
nearest-neighbor coupling only.

The precise setup to observe the quantum phases experimentally depends on
the realization of the dipolar particles. However, here we describe some
of the basic characteristics of the phases which could be detected.  The
coupling of the tubes in the SLL phase can distinguish it from the
Luttinger phase of decoupled tubes. This coupling could, e.g., be
detected by exciting the dipole mode for part of the tubes and detecting
the induced center of mass momentum of the remaining tubes. Further the
frequency of the dipolar mode can give information on the state inside the
tubes~\cite{PedriChiofalo2007}. The stripe and checkerboard orders show characteristic
density-density correlations which could be
detected measuring the noise-correlation spectrum of time-of-flight
images~\cite{AltmanLukin2004}. Before performing the time-of-flight
measurement molecules could be dissociated while freezing their position
by an additional strong optical lattice.

We would like to thank H.-P.~B\"uchler, M.~K\"ohl, O.~Parcollet, P.~Pedri,
and G.~Shlyapnikov for fruitful discussions. This work was partly
supported by the SNF under MaNEP and Division II, and by the
U.~S.~Department of Energy, Office of Science, under Contract
No.~DE-FG02-07ER46424. Furthermore, we would like to thank the Institute
Henri Poincare -- Centre Emile Borel for its hospitality. CK acknowledges
support of the CNRS and of the RTRA network `Triangle de la Physique'.



\end{document}